\documentclass[a4paper,fleqn,usenatbib]{mnras}

\usepackage{newtxtext,newtxmath}

\usepackage[T1]{fontenc}
\usepackage{ae,aecompl}

\usepackage{graphicx}	
\usepackage{amsmath}	
\usepackage{amssymb}	

\title[Inversions of the Ledoux discriminant]{Inversions of the Ledoux discriminant: a closer look at the tachocline}

\author[G. Buldgen et al.]{
Ga\"el Buldgen$^{1}$\thanks{E-mail: gbuldgen@ulg.ac.be},
 S. J. A. J. Salmon$^{1}$,
 M. Godart$^{1}$,
 A. Noels$^{1}$,
 R. Scuflaire$^{1}$,\and and
 M. A. Dupret$^{1}$, 
 D. R. Reese$^{2}$,
 J. Colgan$^{3}$, 
 C. J. Fontes$^{3}$,
 P. Eggenberger$^{4}$,\and and
 P. Hakel$^{3}$, 
 D. P. Kilcrease$^{3}$, 
 O. Richard$^{5}$
\\
$^{1}$STAR Institute, Universit\'e de Li\`ege, All\'ee du Six Ao\^ut 19C, B-4000 Li\`ege, Belgium\\
$^{2}$LESIA, Observatoire de Paris, PSL Research University, CNRS, Sorbonne Universit\'e, UPMC Univ. Paris 06, Univ. Paris Diderot,\\ Sorbonne Paris Cit\'e, 5 place Jules Janssen, 92195 Meudon Cedex, France \\
$^{3}$ Los Alamos National Laboratory, Los Alamos, NM 87545, USA \\
$^{4}$ Observatoire de Gen\`eve, Universit\'e de Gen\`eve, 51 Ch. Des Maillettes, CH$-$1290 Sauverny, Suisse \\
$^{5}$ LUPM, Universit\'e de Montpellier, CNRS, Place E. Bataillon, 34095 Montpellier Cedex, France 
}

\date{Accepted XXX. Received YYY; in original form ZZZ}

\pubyear{2017}

\begin{document}
\label{firstpage}
\pagerange{\pageref{firstpage}--\pageref{lastpage}}
\maketitle

\begin{abstract}
Modelling the base of the solar convective envelope is a tedious problem. Since the first rotation inversions, solar modellers are confronted with the fact that a region of very limited extent has an enormous physical impact on the Sun. Indeed, it is the transition region from differential to solid body rotation, the tachocline, which furthermore is influenced by turbulence and is also supposed to be the seat of the solar magnetic dynamo. Moreover, solar models show significant disagreement with the sound speed profile in this region. In this paper, we show how helioseismology can provide further constraints on this region by carrying out an inversion of the Ledoux discriminant. We compare these inversions for Standard Solar Models built using various opacity tables and chemical abundances and discuss the origins of the discrepancies between Solar Models and the Sun. 
\end{abstract}




\section{Introduction}\label{SecIntroA}

	The base of the solar convective envelope has always been under the spotlight and solar scientists are still debating today on the implications of the tachocline modelling on the magnetic and rotational properties of the Sun. Historically, the tachocline was defined by Jean-Paul Zahn and Edward Spiegel in $1992$ as the transition region between the differentially rotating convective envelope and the uniformly rotating radiative zone. It came in the spotlight after the successful inversions of the rotation profile of the Sun \citep{Kosovichev88Rota}, revealing the importance of this very thin region, extending over less than $4\%$ of the solar radius \citep{corbard99,Elliott99Tacho} between $0.708$ and $0.713$ $r/R_{\odot}$. Historically, solar rotation was, and still is, an issue. The transition in the rotation rate is thought to result from the combined effects of multiple complex physical processes \citep[see][for example]{SpiegelZahn1992, KumarWaves, Spruit99}. The tachocline encompasses all uncertain phenomena of stellar interiors \citep[see][and references therein]{Hughes2007}: it is influenced by turbulent convection, rotational transport and is thought to be the seat of the solar dynamo. It is strongly impacted by convective penetration \citep[see][for a recent study]{JCD11Overshoot} and therefore is influenced by the various studies aiming at refining the current mixing-length formalism used for stellar convection \citep{ZahnMixLength, ZhangTurbConvI, ZhangTurbConvII, Zhang13NonLoc}.  Thus, this very thin region of the solar structure materializes all weaknesses of standard models. 
	
	Besides convective penetration and horizontal turbulence, the magnetic properties of the solar wind also play a strong role, since the extraction of angular momentum will depend on the surface conditions to properly reproduce the current solar rotation profile  \citep[see][for a complete review]{Hughes2007}. In recent years, the advent of 3D hydrodynamical simulations has enabled more in-depth studies of the solar tachocline \citep{Garaud2002I, Garaud2008TachoII, Acevedo2013TachoIII}. However, while such studies are crucial to gaining more knowledge on the potential interactions between the various processes acting in this region, they often operate at a turbulence regime far lower than what is expected at the base of the solar convection zone and do not include all physical processes potentially acting in these regions. Moreover, there would still be a long way to go from a perfect depiction of the current state of the tachocline in a numerical simulation, to the inclusion of its effects in a stellar evolutionary code \citep{Brun3D}. 

	While its effects have first been seen in the rotation profile of the Sun, the tachocline also leaves a mark in the sound speed profile of the Sun \citep{MonteiroOne,AntiaChitre98}. The effects of extra mixing acting below the convective zone during the evolution of the Sun are also witnessed in the lithium and beryllium abundances, whose depletion is closely related to the amount of mixing required. Some studies in global helioseismology have attempted to include its effects in their modelling and determine the changes in the sound speed profile and helium abundance resulting from its presence \citep{Richard96Sun,Richard97Sun,Brun02TachoMix,Richard2004DiffCirc}. With this study, we show that a more efficient thermodynamical quantity can be determined to constrain the stratification near the tachocline. This inversion involves the Ledoux discriminant\footnote{In some references, this quantity can be denoted as the Ledoux convective parameter or convective parameter.}, defined as $A=\frac{d \ln \rho}{dr}-\frac{1}{\Gamma_{1}}\frac{d \ln P}{dr}$, with $r$ the radial position, $P$ the pressure, $\rho$ the density and $\Gamma_{1}= \frac{\partial \ln P}{\partial \ln \rho} \vert_{S}$, the first adiabatic exponent, with $S$ the entropy. This quantity is related to the Brunt-V\"ais\"al\"a frequency and is consequently an excellent probe of local variations near the base of the convective zone.
	
	Inversions of the Ledoux discriminant have been performed by \cite{GoughPmod}, who attempted to generalize its use to stars other than the Sun \citep{GoughAster, TakataWhite}. Unfortunately, the diagnostic potential of this inversion was never used to calibrate the stratification just below the convective zone. Moreover, no comparison between Standard Solar Models (SSMs) with various abundances has been performed, although this inversion probes the region were the low metallicity models seem to be at odds with solar structure \citep{SerenelliComp}. In the following sections, we carry out inversions for SSMs built with various abundances and opacity tables. We illustrate the importance of additional mixing below the limit of the adiabatic convection zone at $0.713r/R_{\odot}$ \citep{KosConv} to reduce inaccuracies between SSMs and the Sun. We discuss how this inversion, coupled with classical helioseismic tools can be used to carry out an in-depth study of the solar tachocline.

\section{Inversion of the Ledoux discriminant for standard solar models}\label{secASSM}

The SSMs used in this study are built using CLES, the Li\`ege stellar evolution code \citep{ScuflaireCles}, for which the frequencies were computed with LOSC, the Li\`ege oscillation code \citep{ScuflaireOsc}. All models used the Free equation of state \citep{Irwin} and either the OPAL \citep{OPAL}, or OPLIB \citep{Colgan} opacity tables. We considered the AGSS09 \citep{AGSS09} and GN93 \citep{GrevNoels} heavy elements abundances. 

The structural kernels used are those of the $\left( A, \Gamma_{1} \right)$ structural pair \citep{GoughPmod}. Both structural kernels and inversions were computed using a customized version of the InversionKit software using the SOLA inversion technique \citep{Pijpers}. We used the solar seismic dataset of \cite{BasuSun} and followed their definitions of the error bars for the SOLA method. 
 
\subsection{Analysis of the contributions to the variations of $A$}
\begin{figure}
	\centering
		\includegraphics[width=8cm]{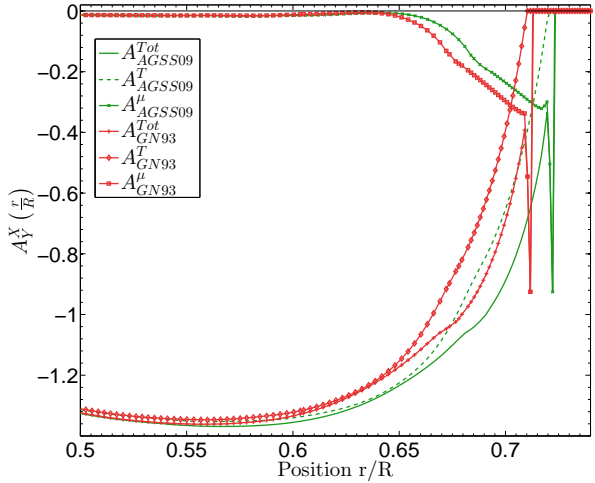}
	\caption{Decomposition of the Ledoux discriminants in its contribution from temperature and mean molecular weight gradients. }
		\label{figADecomp}
\end{figure}

Before showing actual inversion results, we analyse in a more detailed way the separate contributions to the $A$ profile coming from chemical composition and temperature gradients, with the help of the equation of state 
\begin{align}
A=-\frac{r \delta}{H_{p}}\left( \nabla_{ad}-\nabla  + \frac{\phi}{\delta}\nabla_{\mu}\right),
\end{align}
where $H_{P}$ is the pressure scale height, $T$ the temperature, $\mu$ the mean molecular weight, $\nabla_{ad}=\left(\frac{\partial \ln T}{\partial \ln P}\right)_{S}$, $\nabla=\frac{\partial \ln T}{\partial \ln P}$, $\nabla_{\mu}=\frac{\partial \ln \mu}{\partial \ln P}$, $\delta=-\frac{\partial \ln \rho}{\partial \ln T}$, $\phi=\frac{\partial \ln \rho }{\partial \ln \mu}$. The following quantities are illustrated in figure \ref{figADecomp} for AGSS$09$ and GN$93$ SSMs built using the OPLIB opacities
\begin{align}
A^{T}&=-\frac{r \delta}{H_{p}}\left( \nabla_{ad}-\nabla \right), \\
A^{\mu}&=-\frac{r \phi}{H_{p}}\nabla_{\mu}.
\end{align}

In figure \ref{figADecomp}, we see that the steepness of the temperature gradient changes significantly when using different abundance tables. However, the contribution of the mean molecular weight gradient term, $A^{\mu}$, is rather small and similar for both models, except in a region which extends approximately $0.03$ solar radii just below the base of the convective zone, in the tachocline. Below this very localized region, the contribution of the term $A^{T}$ quickly takes over and dominates the behaviour of $A$, until one reaches the region influenced by thermonuclear reactions.

\subsection{Verification of the reliability of the inversion}

\begin{figure}
	\centering
		\includegraphics[width=8.5cm]{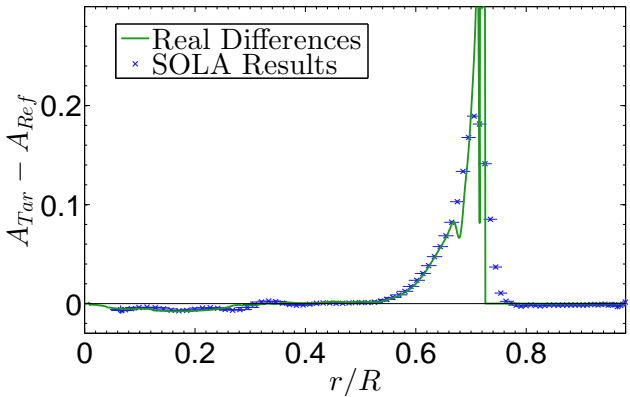}
	\caption{Illustration of the hare-and-hounds exercises performed for SSMs using the SOLA method. The actual differences are given in green while the inversion results are presented with their error bars in blue.}
		\label{figHH}
\end{figure}

	 In figure \ref{figHH}, we illustrate the results of hare-and-hounds exercises between SSMs using the same mode set of frequencies as those observed in the Sun and their actual uncertainties from the observations. The hare is a SSM built using the OPLIB opacities, the GN$93$ abundances and the Opal equation of state \citep{Rogerseos}, it is denoted as $A_{Tar}$ in figure \ref{figHH}, while the hound, denoted as $A_{Ref}$ is a SSM built using the AGSS$09$ abundances, the OPLIB opacities and the FreeEOS equation of state. We see that most of the profile is well reproduced, with the exception of the steep localized features between $0.7$ and $0.713$ $r/R_{\odot}$. This is due to the finite width of the averaging kernels which cannot resolve such steep variations. Similar problems have been presented in previous exercises \citep{KosovReview}. We also checked the averaging kernels \citep[see][for details]{Pijpers} of the inversion to further ensure reliability. From its definition, we see that $A$ is of very low amplitude in most of the convective zone. However, as seen in figure \ref{figASSMpapier}, the inversion sometimes attributes an unrealistically large non-zero value to the $A$ differences in such regions, which is clearly an artefact. It does not seem that this non-zero value stems from an inaccurate fit of the target. Instead, it seems to be linked to the amplitude of the lobe just below the convective zone in the inversion, since models built with the GN$93$ abundances show a better agreement with the expected very low value than the AGSS$09$ models. Similar problems have also been seen in \cite{KosovReview} for solar inversions, but they do not affect the diagnostic in the radiative region.

\subsection{Inversion results}

\begin{figure*}
	\centering
		\includegraphics[width=13.8cm]{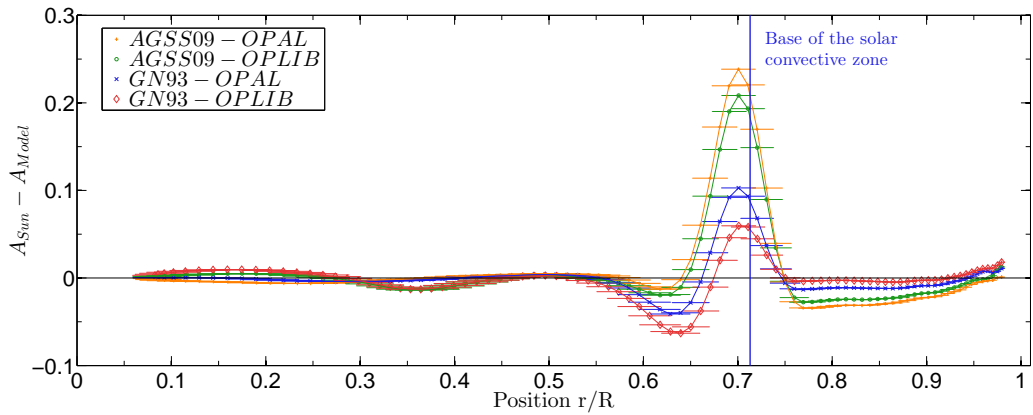}
	\caption{Effects of the opacities on the Ledoux discriminant profile of SSMs. The orange and green symbols are related to SSMs built with the AGSS$09$ abundances and the OPAL and OPLIB opacity tables, respectively. The red and blue symbols are related to the GN$93$ SSMs. Horizontal error bars show the interquartile width of the averaging kernels while the vertical error bars are the 1$\sigma$ errors from the propagation of the observational 1$\sigma$ errors.}
		\label{figASSMpapier}
\end{figure*}

In figure \ref{figASSMpapier}, we illustrate inversions of the Ledoux discriminant for SSMs using both the AGSS09 and GN93 abundances alongside the OPAL and OPLIB opacities. Just below the convective zone, the GN93 models, in red and blue, seem to be the best in terms of Ledoux discriminant inversions while the AGSS09 models, in orange and green, show large discrepancies. This result is not surprising since models built with the AGSS09 abundances do not set the base of the convective zone at its location determined from helioseismology, and the sound speed inversion of these models also show a strong disagreement in this region. 

However, it is more surprising to see that in the region just below the tachocline, around $0.65$ normalized radii, the GN$93$ models seem to perform less well than the AGSS$09$ models. This trend is present for both the OPAL and the OPLIB opacity tables. Furthermore, hare-ands-hounds exercises have confirmed that this trend is not affected by the artefact observed in the convective zone. From figure \ref{figADecomp}, we can see that the mean molecular weight gradient term cannot be held responsible for the deviation seen between $0.6$ and $0.65$ $r/R_{\odot}$, but rather that it stems from a too steep temperature gradient in this region, which is either a consequence of the higher heavy elements mass fraction of this model or the opacity profile in this region. 

As shown in the next section, including some mixing below the convection zone can reduce the discrepancies, but the most straightforward way to correct them is to lower the metallicity of the SSMs. However this would imply acknowledging a significant increase in the inaccuracies of the layers just below the convection zone, as can be seen from inversions of AGSS09 models in figure \ref{figASSMpapier}.

The overestimation of the temperature gradient around $0.65$ $r/R_{\odot}$ in SSMs built using the GN$93$ abundances would also imply that their agreement with the Sun results, at least partially, from a certain degree of compensation. The steeper temperature gradient, resulting from the higher metallicity of these models, reduces the discrepancies observed in the tachocline. This compensation would also explain why the GN$93$ models reproduce well the steep variations in sound speed observed at $0.713$ $r/R_{\odot}$, resulting from the transition from the adiabatic to the radiative temperature gradient in the Sun without the need for any convective penetration.

From a theoretical perspective, the Schwarzschild limit derived by using the mixing-length theory, as done in SSMs, should not be located at the same depth as the observed solar transition. In the Sun, the variation in the sound speed is located at the kinetic limit of the convective elements which is influenced by convective penetration and differs from the limit derived solely by the mixing-length theory in SSMs. Moreover, the effects of shear could induce an additional mixing of the chemical elements below the transition in temperature gradients, meaning that the fully mixed zone will extend slightly deeper than $0.713$ $r/R_{\odot}$. This justifies investigations of the impact of an additional chemical mixing in the regions just below the transition of the temperature gradients.

\section{Impact of additional mixing on the Ledoux discriminant profile}

From the test cases of the Sect. \ref{secASSM}, it seems that most of the discrepancies in the SSMs are stemming from a very narrow region just below the convective zone. From the chemical point of view, a certain degree of mixing is expected in the tachocline, due to shear, horizontal shear-generated turbulence and convective plumes. However, the temperature profile in this region is not theoretically constrained. Various studies used phase shifts \citep{RoxVor94PhaseShift} of solar oscillations to constrain the transition in the temperature gradients, finding that while the transition is not discontinuous, it should still be quite steep \citep{MonteiroOne}. 

\begin{figure*}
	\centering
		\includegraphics[width=13.8cm]{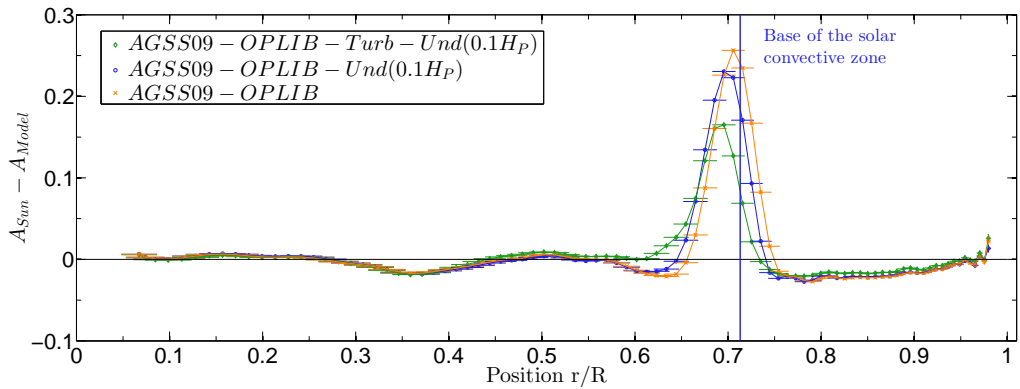}
	\caption{Inversions of the Ledoux discriminant for models built with the FreeEOS equation of state, the OPLIB opacities and the AGSS$09$ abundances including extra mixing. The orange symbols are associated with the SSM presented in figure \ref{figASSMpapier}, the blue symbols are associated with a model including a small undershooting coefficients. The green symbols are associated with a model including a small convective penetration and an additional diffusive mixing localised below the transition in the temperature gradient. The horizontal and vertical error bars follow the definition adopted in previous figures.}
		\label{figAMix}
\end{figure*}

Using Ledoux discriminant inversions, we have analysed the impact of adding turbulent diffusion below the convective zone. We used solar models including the AGSS$09$ abundances, the OPLIB opacities and the FreeEOS equation of state. All our non-standard models included a $0.1H_{p}$ convective penetration at the base of the envelope, assuming full chemical mixing, so that the temperature gradients transition would fit that of the Sun. To approximate the effects of a tachocline in one of our models, we used an exponential formulation of the diffusion coefficient with a scale height related to density, as done in \cite{Miglio07}. The extent of the mixing is limited to a small zone of around $0.05$ solar radii, of the order of the radial extent of the tachocline \cite{corbard99}. The inversion results of the Ledoux discriminant are illustrated in figure \ref{figAMix}. As seen from the green symbols when compared to the orange symbols, associated with the AGSS09 SSM, the discrepancies with the Sun are reduced by around $0.08$ due to the extra mixing processes below the convective zone. However, the sound speed profile is not significantly improved by these changes. 

From these tests, we conclude that mixing alone cannot correct the observed discrepancies and that the remaining errors may stem from inaccuracies in the temperature profile. This is confirmed by the fact that further increasing the intensity and extent of the mixing results in similar discrepancies than with a SSM. The observed inaccuracies could stem from various effects, such as for example the surrounding opacity profile or the modelling of the transition of temperature gradients in the tachocline. Indeed, there is currently no agreement on the description of the transition from the adiabatic to radiative gradient around $0.713r/R_{\odot}$. Ad-hoc parametrizations could be used to analyze the effects of attempting to properly reproduce the transition region in the sound-speed and Ledoux discriminant profiles.  

In addition to these processes, additional refinements of the SSMs might be required to reproduce the solar structure. For example, using individual elements instead of an average metallic element might induce changes in the opacity calculations. Moreover, using individual elements would allow a better inclusion of partial ionization effects in the computation of the diffusion velocities. Such refinements would impact both temperature and chemical composition gradients. Moreover, additional updates on the equation of state might slightly alter the results and while it would not eliminate the errors just below the convection zone, it might influence the deeper radiative regions. The link between such effects and the lithium abundances should also be carefully studied with models including the effects of rotation and properly reproducing the lithium and berylium abundances. 

\section{Conclusion}

Inversions of the Ledoux discriminant offer very strong constraints on the stratification just below the solar convective zone. Thanks to its very localized sensitivity, it can be used to efficiently constrain both temperature and mean molecular weight gradients in the solar tachocline. In SSMs, the Ledoux discriminant inversion can lead to interpret the reason of the success of the high metallicity models, as the result of a form of compensatory behaviour in the temperature gradient near the base of the convective envelope. Combining the knowledge of these diagnostics, we are already able to demonstrate that a certain amount of diffusive mixing below the temperature gradient transition at $0.713$ $r/R_{\odot}$ helps in reducing the disagreement of low metallicity models with the Sun. However, additional mixing due to the tachocline is insufficient to eliminate the observed discrepancies and the remaining inaccuracies are the proof of an inadequate reproduction of the temperature gradient in the lower part of the tachocline. Parametrizations of the transition region could provide further insights into the problem, keeping in mind that non-radial variations can also be expected in the tachocline and would thus limit the predictive power of $1D$ structural inversions. Moreover, as a consequence of the current debate in the opacity community, these studies should not aim at providing a full agreement, but rather be used to reinforce the strong links between stellar models, hydrodynamical simulations and theoretical developments, for progress in these fields is necessary to improve our current depiction of solar and therefore stellar interiors.




\bibliographystyle{mnras}
\bibliography{biblioarticleA} 

\bsp	
\label{lastpage}
\end{document}